\documentclass[a4paper,12pt]{article}
\usepackage[utf8]{inputenc}
\usepackage{authblk}
\usepackage{amsmath}
\usepackage{amssymb}
\usepackage{amsfonts}
\usepackage{array}
\usepackage{mathtools}
\usepackage[square,numbers]{natbib}
\usepackage{lmodern,pdftexcmds}

\newtheorem{assumption}{Assumption}

\usepackage[hidelinks]{hyperref}
\hypersetup{colorlinks=true,
  linkcolor=blue,urlcolor=black,citecolor=blue}

\usepackage{cleveref}
\usepackage{comment}
\usepackage{color}

\crefname{assumption}{Assumption}{Assumptions}
\newcommand{\del}{\ensuremath{\partial}}

\usepackage[normalem]{ulem}
\usepackage{soul}

\title{Emergence of coupled Korteweg–de Vries equations in $m$ fields} %% Article title
\author{Sharath Jose, Manas Kulkarni, Vishal Vasan}
\affil{International Centre for Theoretical Sciences, Tata Institute of Fundamental Research, Bangalore 560089, India}
\date{}
\begin{document}

\maketitle




%% Abstract
\begin{abstract}
The Korteweg–de Vries (KdV) equation is of fundamental importance in a wide range of subjects with generalization to multi-component systems relevant for multi-species fluids and cold atomic mixtures. We present a general framework in which a family of multi-component KdV (mKdV) equations naturally arises from a broader mathematical structure under reasonable assumptions  on the nature of the nonlinear couplings. In particular, we derive a universal form for such a system of $m$ KdV equations that is parameterized by $m$ non-zero real numbers and two symmetric functions of those $m$ numbers. Secondly, we show that physically relevant setups such as $N\geq m+1$ multi-component nonlinear Schrödinger equations (MNLS), under scaling and perturbative treatment, reduce to such a mKdV equation for a specific choice of the symmetric functions. The reduction from MNLS to mKdV requires one to be in a suitable parameter regime where the associated sound speeds are repeated.  Hence, we connect the assumptions made in the derivation of mKdV system to physically interpretable assumptions for the MNLS equation. Lastly, our approach provides a systematic foundation for facilitating a natural emergence of multi-component partial differential equations starting from a general mathematical structure.
\end{abstract}

%% Keywords
%\begin{keyword}
%\end{keyword}




\tableofcontents 

\section{Introduction}
\label{sec:intro}
%The Korteweg--de Vries (KdV) equation offers a framework within which nonlinear phenomena in several systems can be described. 

The Korteweg--de Vries (KdV) equation is one of the most widely studied partial differential equations in literature~\cite{Ablowitz_Segur_1981,Novikov_etal_1984,Ablowitz_2011}. It finds relevance in a wide range of fields including, but not limited to, fluid dynamics \citep{Wijngaarden_1968JFM,Hammack_Segur_1974JFM}, low-dimensional classical systems \citep{Zabusky_Kruskal_1965PRL} and nonlinear optics~\cite{Kivshar_1993JQE}. Notable among its properties are an infinite number of conserved quantities \citep{Miuraetal_1968JMP,Miura_1976SIAMRev}, its Hamiltonian structure \citep{Gardner_1971JMP,Zakharov_Faddeev_1971FAA} and the existence of multi-soliton solutions \citep{Hirota_1971PRL,Hirota_2004} that interact cleanly without the emission of dispersive waves. 
These solitons emerge in a variety of physical contexts, from water waves to Bose-Einstein condensates. Understanding the KdV equation has influenced developments in mathematical physics, nonlinear dynamics, and even field theories.

There has been a surge of activities in generalising the KdV to multi-component systems. This is important for understanding systems with several species such as multi-component fluids and cold atomic mixtures. When interactions between multiple nonlinear waves become important, a system of coupled KdV equations often ends up being the relevant framework to study the dynamics. Examples include the interactions of nonlinear waves in geophysical settings \citep{Gear_Grimshaw_1984SAM,Majda_Biello_2003JAS,Lou_etal_2006JPAMG} and soliton dynamics in a Bose--Einstein condensate \citep{Brazhnyi_Konotop_2005PRE}. The coupling between the equations could be through a linear term (non-dispersive or dispersive), a nonlinear term or a combination of the two. Depending on the model at hand, a host of solitary wave phenomena can be observed \citep{Grimshaw_Malomed_1994PRL,Champneys_etal_2000PLA,Grimshaw_Iooss_2003MCS,Fochesato_etal_2005PD}.
Motivated by the fact that the KdV equation is integrable, there have been numerous studies devoted to identifying integrable coupled KdV systems \citep{Hirota_Satsuma_1981PL,Satsuma_Hirota_1982JPSJ,Ito_1982PL,Fuchssteiner_1982PTP,Baker_etal_1995PLA,Karasu_1997JMP,Sakovich_1999JNMP,Foursov_2000IP,Li_2004JPSJ,Wang_2010AMC}. Such systems have also served as a testbed for different techniques geared towards finding analytical nonlinear solutions \citep{Yan_2003CSF,Ganji_Rafei_2006PLA,Hu_etal_2006PLA,Wazwaz_2011CEJP}. Barring a few exceptions \citep{Svinolupov_1991,Gurses_Karasu_1996PLA,Miller_Christiansen_2000PS}, it is to be noted that most studies on coupled KdV systems are restricted to those involving only two or three fields.

Often, how multi-component KdV (mKdV) systems naturally emerge from a general mathematical structure is much less understood. In this work, we present a framework for obtaining coupled KdV systems of $m\geq 2$ fields starting from a general 
Hamiltonian picture and making reasonable assumptions on the nature of the coupling between the fields. 
We then show that the obtained mKdV equations naturally occurs in a cold atomic gas of $N\geq m+1$ species
described by a multi-component nonlinear Schr\"odinger equation (MNLS). Using perturbative techniques, the MNLS system dynamics can be mapped to that of a system of KdV equations. The sound speeds of the system play a vital role in determining the perturbative dynamics. If the sound speeds are distinct, a clear separation of time scales exists that allows for the system of KdV equations to decouple \citep{Swarup_etal_2020JPA,Nimiwal_etal_2021JPA}. 
However, when the sounds speeds are repeated, we show here that the relevant dynamics is described by a set of $m$ coupled KdV equations.

The paper is structured as follows. In section~\ref{sec:mkdv_main}, we discuss the emergence of multi-component KdV equations starting from a generic mathematical structure. In section~\ref{sec:mkdv}, we provide a physical realization of such mKdV equations from a  multi-component nonlinear Schr\"odinger equation of $N$ species subject to scaling and perturbative analysis. We provide conclusions along with an outlook in section~\ref{sec:conc}.

\section{Multi-component KdV}
\label{sec:mkdv_main}
Consider the multi-component KdV type equation
\begin{align}
\partial_t \frac{\delta \mathcal P}{\delta u_j} = \partial_x \frac{\delta \mathcal H}{\delta u_j}\label{eqn:mKdV}
\end{align}
which describes the coupled evolution of $m$ fields $u_j(x,t)$. Here 
\begin{align}
    P_m=\int \mathcal P(u_1,\ldots,u_m)~dx \, ,\quad H_m =  \int \mathcal H(u_1,u_2,\ldots,u_m)~dx \label{eqn:PH_m}
\end{align}
are the momentum and Hamiltonian with densities $\mathcal P,\mathcal H$ respectively. We suppose the density $\mathcal P$ is a quadratic expression, \emph{i.e.} 
\begin{align}
\mathcal P(u_1,\ldots,u_m) = \frac  1 2 \sum_{kj}L_{kj}u_k u_j\label{eqn:Pm:quad-form}
\end{align}
where $L_{kj}$ are elements of a symmetric invertible matrix. As a consequence, the equation for evolution of the $m$ fields is of the form
\begin{align}
\partial_t u = L^{-1} \partial_x \frac{\delta \mathcal H}{\delta u}\label{eqn:mKdV:GenHamil}
%\label{eq:dtu}
\end{align}
where $u=[u_1,\ldots,u_m]^T$ denotes the vector of fields and $\frac{\delta \cdot}{\delta u}$ is the variation with respect to each of the fields $u_j$, resulting in a $m-$vector of fields. An immediate consequence of this form is that both $P_m$ and $H_m$ are conserved quantities for \cref{eqn:mKdV} as we now show. Indeed, consider the rate of change of $H_m$
\begin{align}
\frac{d}{dt} H_m = \int \sum_j \frac{\delta \mathcal H}{\delta u_j}\sum_{k}L^{-1}_{kj}\partial_x \frac{\delta \mathcal H}{\delta u_k}~dx  = \sum_{kj}\int L^{-1}_{kj}\frac{\delta \mathcal H}{\delta u_j}\partial_x \frac{\delta \mathcal H}{\delta u_k}~dx
\end{align}
Since $L^{-1}_{kj}=L^{-1}_{jk}$, each term in the summation can be written as an exact derivative. Hence the expression above vanishes and $H_m$ is conserved. Similarly, consider the rate of change of $P_m$
\begin{align}
\frac{d}{dt}P_m = \int \sum_j \frac{\delta \mathcal P}{\delta u_j}\sum_{k}L^{-1}_{kj}\partial_x \frac{\delta\mathcal H}{\delta u_k}~dx = \sum_{kj}\sum_i \int L_{ij}L^{-1}_{kj}u_i\partial_x \frac{\delta \mathcal H}{\delta u_k}~dx 
\end{align}
However note 
\begin{align}
\sum_{j}L_{ij}L^{-1}_{kj} = \delta_{ik},\quad \mbox{the }m\times m\mbox{ identity matrix} 
\end{align}
and hence
\begin{align}
\frac{d}{dt}P_m =  \sum_k \int u_k \partial_x\frac{\delta \mathcal H}{\delta u_k}~dx = - \sum_k \int \frac{\delta\mathcal H}{\delta u_k}\partial_x u_k~dx = 0 
\end{align}
since the integrand in the latter expression is an exact derivative (for each $k$).

We emphasize the form of the Hamiltonian is irrelevant to the conservation of both the total momentum and the Hamiltonian; a symmetric invertible matrix $L$ in \cref{eqn:Pm:quad-form} is sufficient. The form of the Hamiltonian is however relevant when we want to specify the kinds of coupling present in the coupled KdV model. The specific form of the Hamiltonian considered in this work is given by 
\begin{align}
\mathcal H (u_1,\ldots,u_m) = \frac 1 2 \sum_{kj}L_{kj}~\partial_x u_k~ \partial_x u_j + \sum_{ijk} R_{ijk}u_iu_ju_k\label{eqn:Hm}
\end{align}
This ensures the fields $u_j$ are linearly decoupled. The cubic term above ensures we obtain quadratic advective nonlinearities as one expects for KdV-type equations. We also employ the shorthand for the nonlinear coupling term
\begin{align}
    \mathcal R = \sum_{ijk}R_{ijk}u_iu_ju_k\label{eqn:R}
\end{align}
and hence, using \cref{eqn:mKdV:GenHamil}, we obtain the following form of the mKdV system
\begin{align}
\partial_t u + \partial_x^3 u - L^{-1}\partial_x \frac{\delta \mathcal R}{\delta u} = 0\label{eqn:mKdV2}
\end{align}
We re-write \cref{eqn:mKdV2} in the following equivalent form for a specific field $u_j$
\begin{align}
\partial_t u_j + 6u_j\partial_x u_j + \partial_x^3 u_j =  \partial_x \sum_{k}\left(L^{-1}_{kj} \frac{\delta \mathcal R}{\delta u_k} +3\delta_{kj}u_k^2\right)\label{eqn:mKdV3}
\end{align}
The above form emphasizes the fact that \cref{eqn:mKdV2} is a system of $m-$coupled KdV fields where the coupling entirely arises from the nonlinear term $\mathcal R$ in the Hamiltonian \cref{eqn:Hm}.

\subsection{The universal form of all-to-all and equal coupling}\label{subsec:universal}
The right-hand side of \cref{eqn:mKdV3} consists entirely of terms quadratic in $u_k$. The most general form of such an equation is given by
\begin{align}
\partial_t u_j + 6u_j\partial_x u_j + \partial_x^3 u_j = 3\partial_x 
\sum_{ln}N_{jln} u_lu_n
\label{eqn:mKdV4}
\end{align}
where $N_{jln}$ are real-numbers representing the coupling coefficients. Note the coefficient of a particular term $u_lu_n$ on the right-hand side of \cref{eqn:mKdV4} is $N_{jln}+N_{jnl}$. In particular the individual term $N_{jln}$ does not appear on its own. This motivates the following assumption.
{\assumption{$N_{ipq}=N_{iqp}$ for every $i,p,q$\label{ass:Q}\\ }}

\noindent A very similar argument also applies to \cref{eqn:mKdV3} where once again we expect only permutation invariant quantities to appear. Indeed note 
\begin{align}
\frac{\delta\mathcal R}{\delta u_p} &=  \sum_{ij}( R_{ijp}+R_{ipj}+R_{pij} )u_iu_j\:
\end{align}
and hence the coefficient of a specific term $u_iu_j$ on the right-hand side  of \cref{eqn:mKdV3} is given by
\begin{align}
R_{ijp}+R_{ipj}+R_{pij} + R_{jip}+R_{jpi}+R_{pji}\:
\end{align}
Hence, we have the following assumption on the nonlinear couplings in the Hamiltonian.
{\assumption{
$R_{jpq}$ is invariant under every permutation of $j,p,q$ \label{ass:R} \\}}    

\noindent Note that \cref{ass:Q,ass:R} are both natural and do not restrict the class of models considered. On the other hand, as will be seen in the following, the next assumption has consequences on the structure of the equations. The following assumption encodes the fact that all fields $u_j$ are coupled to one another in precisely the same manner, \emph{i.e.} the coupling is all-to-all and equal.

{\assumption{$N_{jpq}=N_{pq}$, or in other words $N$ is independent of the first index.\label{ass:all2all}\\}}

We now show that the multi-component coupled system of KdV-type stated in \cref{eqn:mKdV} corresponding to total momentum given by \cref{eqn:Pm:quad-form} and Hamiltonian given by \cref{eqn:Hm}, under \crefrange{ass:Q}{ass:all2all}, is of a universal form. In particular we prove that the coupling constants must be of the form
\begin{align} \label{eqn:Nuniversal}
       N_{pq} &= s_1(\bar w)w_qw_p + s_2(\bar w)w_q\delta_{pq}\\
       L_{ij} &= s_2(\bar w)w_iw_j +  \left(1-s_2(\bar w)\sum_iw_i\right)w_j \delta_{ij}\label{eqn:Luniversal}\\
       R_{pqj} &= s_1(\bar w)w_pw_qw_j - \left(1-s_2(\bar w)\sum_iw_i\right)w_j\delta_{pq}\delta_{qj} \label{eqn:Runiversal}
\end{align}
where $w_i$ are $m$ real non-zero numbers and $s_1,s_2$ are arbitrary symmetric functions of the $w_i$ such that $s_2(\bar{w}) = s_2(w_1,w_2,\ldots,w_m)\neq \left(\sum w_j\right)^{-1}$. Moreover the matrix $L$ is symmetric and invertible. As a consequence, the total momentum and Hamiltonian are conserved for this universal all-to-all and equally coupled KdV system. Note that $L$ and $R$ are only defined up to an overall scale factor since multiplying the total momentum $P_m$ and the  Hamiltonian $H_m$ by a scalar does not change the equations for the fields $u_j$.

We begin deriving the universal form by equating the right-hand sides of  \cref{eqn:mKdV3} and \cref{eqn:mKdV4} to obtain
\begin{align}
%\[
\sum_{kln}L^{-1}_{kj} ( R_{lnk}+R_{lkn}+R_{kln} )u_lu_n +3\sum_{k}\delta_{kj}u_k^2 = 3\sum_{ln}N_{jln} u_lu_n %\]
\end{align}
Since the above equation is valid for every set of $m-$fields $u_j$, we differentiate both sides first with respect $u_p$, and then with respect to $u_q$ to find 
\begin{align}
\sum_{kln}L^{-1}_{kj} ( R_{lnk}+R_{lkn}+R_{kln} )(\delta_{lp}\delta_{nq}  + \delta_{np}\delta_{lq}) +6\sum_{k}\delta_{kj}\delta_{kp}\delta_{kq} \nonumber \\
\quad\quad = 3\sum_{ln}N_{jln} (\delta_{lp}\delta_{nq} + \delta_{np}\delta_{lq})
\end{align}
which upon employing \crefrange{ass:Q}{ass:all2all} simplifies to the following relationship between the various coupling coefficients
\begin{align}
\delta_{pj}\delta_{qj} + \sum_{k}L^{-1}_{kj}R_{pqk} =  N_{pq}\label{eqn:CCrelation}
\end{align}
Multiplying \cref{eqn:CCrelation} by $L_{ij}$ and summing over $j$, we find (after replacing $i$ with $j$) the following equivalent relation between the coupling coefficients
\begin{align}
R_{pqj} = N_{pq}\sum_i L_{ij} - L_{pj}\delta_{pq} \label{eqn:CCrelation2}
\end{align}
Let us define $w_j = \sum_i L_{ij}$ and suppose that $p\neq q\neq j$. Then from \cref{eqn:CCrelation2} we obtain
\begin{align}
R_{pqj}&=N_{pq}w_j\nonumber\\
\Rightarrow R_{pqj}=s_1(\bar w)w_pw_qw_j\:\mbox{and}&\: N_{pq}=s_1(\bar w)w_pw_q\:,\quad p\neq q\neq j
\label{eqn:RN:off-diag}
\end{align}
where we used the symmetry property of $R_{pqj}$ to conclude the decomposition of $R_{pqj}$ into a tensor product of $w_i$ and $s_1$ is an arbitrary symmetric scalar function of all $w_j$. Hence we obtain the form of the `off-diagonal' coupling constants. 

On the other hand, suppose instead $p\neq q$ and $p=j$. Then from \cref{eqn:CCrelation2} we obtain $R$ for the off-diagonal terms on a face 
\begin{align}
R_{jqj} = N_{jq}w_j \Rightarrow R_{jqj} = s_1(\bar w)w_j^2 w_q\:,\quad q\neq j\label{eqn:R:face}
\end{align}
Now consider the following derived from \cref{eqn:CCrelation2} for $q\neq j$
\begin{align}
       \sum_p R_{pqj} &= w_j \sum_p N_{pq} - \sum_p L_{pj}\delta_{pq}\\ \nonumber
     \Rightarrow   R_{jqj} + R_{qqj} + \sum_p^{\substack{p\neq q\\p\neq j}}R_{pqj} &=  w_j N_{jq} + w_j N_{qq} - L_{qj}
     + w_j \sum_p^{\substack{p\neq q\\p\neq j}}N_{pq}
     \\ \nonumber
     \Rightarrow  R_{jqj} + R_{qqj} + s_1(\bar w)w_jw_q \sum_p^{\substack{p\neq q\\p\neq j}}w_p 
     &=
      s_1(\bar w)w_j^2 w_q + w_j N_{qq} - L_{qj} \\ \nonumber
      &\quad \quad +
      s_1(\bar w)w_jw_q\sum_p^{\substack{p\neq q\\p\neq j}}w_p
      \\ \nonumber
     \Rightarrow s_1(\bar w)w_q^2w_j &= w_j N_{qq} - L_{qj}\\
     \Rightarrow L_{qj}  &= w_j N_{qq} - s_1(\bar w)w_q^2w_j \:,\quad \quad q\neq j \label{eqn:L:offdiag}
\end{align}
where we have used \cref{eqn:R:face} to obtain the off-diagonal form of $L$.

Note that the most general form for $N$ consistent with \cref{eqn:RN:off-diag} is  $N_{qj}=s_1(\bar w)w_qw_j + c_q\delta_{qj}$ where $c_q$ is some function of $q$. From this we immediately obtain  $N_{qq}=s_1(\bar w)w_q^2 + c_q$. Using  the form of $N_{qq}$ in \cref{eqn:L:offdiag} we obtain
\begin{align}
\label{eq:lqj}
%\[
L_{qj} = w_jc_q
%\]
\end{align}
However, since $L_{qj}=L_{jq}$, \cref{eq:lqj} reduces to
\begin{align}
L_{qj} = s_2(\bar w)w_jw_q\quad\mbox{where } c_q = s_2(\bar w)w_q \:, \quad q\neq j
\label{eqn:L:off-diag2}
\end{align}
where $s_2$ is yet another arbitrary symmetric function of all $w_j$. Summarizing the progress so far, we have
\begin{align}
       N_{pq} &= s_1(\bar w)w_qw_p + s_2(\bar w)w_q\delta_{pq}\\
       L_{ij} &= s_2(\bar w)w_iw_j + \hat c_j \delta_{ij}\\
       R_{pqj} &= s_1(\bar w)w_pw_qw_j + \tilde c_j \delta_{pq}\delta_{qj}
\end{align}
where $\hat c_j,\tilde c_j$ are as yet undetermined coupling constants along the diagonal. To determine these, we substitute the above expressions into \cref{eqn:CCrelation2} and find $\hat c_j=-\tilde c_j$. Furthermore
\begin{align}
%\[
w_j = \sum_i L_{ij} \Rightarrow  \hat c_j = \left(1-s_2(\bar w)\sum_iw_i\right)w_j %\]
\end{align}
and thus we obtain the expressions in \crefrange{eqn:Nuniversal}{eqn:Runiversal}.

\subsection*{Invertibility of $L$} It remains to verify that $L$ defined by \cref{eqn:Luniversal} indeed represents an invertible matrix. Recall that we required invertibility to establish conservation of momentum and the Hamiltonian. To check the invertibility of $L$ we investigate its null space. Suppose $v\in\mathbb{R}^n$ is a null-vector of $L$. Then the following holds component-wise for each $i$
\begin{align}
%\[
w_i \left(s_2(\bar w)\sum_j v_jw_j + \left(1-s_2(\bar w)\sum_jw_j\right) v_i\right) = 0%\]
\end{align}
If $s_2(\bar w)=(\sum_j w_j)^{-1}$ then $L$ is a rank-1 matrix and hence not invertible. Suppose now instead that $s_2(\bar w)\neq (\sum_j w_j)^{-1}$. Then $L$ has a null-space if and only if for each $i$
\begin{align}
%\[
w_i = 0 \quad \mbox{OR} \quad v_i = \frac{s_2(\bar w)\sum_jv_jw_j}{s_2(\bar w)\sum_j w_j- 1}%\]
\end{align}
Hence for all $i$ such that $w_i\neq 0$, the corresponding $v_i$ must all be equal to some common $v$ since the right-hand side of the $v_i$ expression above is independent of $i$. But then
\begin{align}
%\[
\forall i \quad v_i = v = \frac{s_2(\bar w)v\sum_jw_j}{s_2(\bar w)\sum_j w_j- 1}\Rightarrow v=0 %\]
\end{align}
This implies that if some $w_i=0$ then the corresponding $v_i$ may be any number and we have a nontrivial null-space (of dimension equal to the number of zero $w_i$). We then conclude that if $w_i\neq 0$ for all $i$ and $s_2(\bar w)\neq (\sum_jw_j)^{-1}$ then $L$ is invertible.

\section{Physical realization of multicomponent KdV}
\label{sec:mkdv}
We slightly shift focus to consider a different coupled-system of (complex-valued) fields, the $N$-field multi-component nonlinear Schr\"{o}dinger equation (MNLS) in 1D
\begin{equation}
    i\frac{\del\psi_k}{\del t}=-\frac{1}{2}\frac{\del^2\psi_k}{\del x^2}+\sum_{j=1}^{N} \alpha_{kj}|\psi_j|^2 \psi_k
  \label{eq:mnls}
\end{equation}
Here $\psi_k$ is a macroscopic wavefunction (with $k=1,\ldots N$) and $\alpha$ is the $N\times N$ real-matrix of coupling constants given by
\begin{align}
  \alpha_{jk} = h + (g_j-h)\delta_{jk} \: \label{eq:alpha}
\end{align}
where $g_j,h$ are positive constants. Note in the cold-atom application, the coupling constants can be accurately tuned to desired values \citep{Inouye_etal_1998,Roati_etal_2007PRL,Thalhammer_etal_2008PRL,Wacker_etal_2015PRA}. The form of the coupling-constants assumed in \cref{eq:alpha} indicates all fields are coupled to one another the strength of the cross-coupling is the same among all fields.

Note that one exact solution (constant plane wave solution) to \cref{eq:mnls} is 
\begin{equation}
    \psi_{0k} = \sqrt{\rho_{0k}}e^{i\theta_{0k}},\quad \text{where}\quad \theta_{0k} =  \overline{\theta}_{0k} - t\sum_{j=1}^{N} \alpha_{kj}|\rho_{0j}|^2
    \label{eq:uni}
\end{equation}
%\sout{It is evident that the uniform state $\psi_k = \sqrt{\rho_{0k}}\; e^{i\theta_{0k}}$ for $\rho_{0k}\geq 0$, is an exact solution to \cref{eq:mnls}. }
It is well-known that the dynamics of small-amplitude perturbations about the uniform state for $N=1$ are governed by the KdV equation. In \citet{Swarup_etal_2020JPA}, the authors showed that a single-scalar KdV equation also captured the dynamics for general $N$, where each field $\psi_k$ was essentially proportional to a solution to KdV.

In this section we show that under suitable conditions on the coupling constants, the dynamics of small-amplitude perturbations to a uniform state are described by a system of coupled KdV fields. This equation is a special case of the general coupled system discussed in \Cref{sec:mkdv_main}. In the following %subsequent section 
we briefly describe the derivation starting from \cref{eq:mnls}, following the ideas laid out in Ref.~\cite{Swarup_etal_2020JPA}.

%\subsection{Hydrodynamic description of MNLS}
\subsection{Nonlinear evolution of perturbations}
With $\rho_k(x,t)$ and $v_k(x,t)$ denoting the density and velocity fields respectively of $k^{\textrm{th}}$ condensate, the Madelung transform 
\begin{align}
%\[
\psi_k(x,t) =\sqrt{\rho_k(x,t)}e^{i\int_{0}^{x}v_k(x',t)dx'} %\]
\end{align}
is used to move to a hydrodynamic framework. Then we have the following equations governing the evolution of $\rho_k(x,t)$ and $v_k(x,t)$ 
\begin{align}
  &\frac{\del \rho_k}{\del t} + \frac{\del}{\del x} \left(\rho_k v_k \right) = 0 \\
  &\frac{\del v_k}{\del t} = -\frac{\del}{\del x}\left[\frac{v_k^2}{2} + \sum_{j=1}^{N}\alpha_{kj}\rho_j - \left(\frac{1}{2}\right)\frac{\del_x^2\sqrt{\rho_k}}{\sqrt{\rho_k}}\right]
\end{align}
A base state with non-negative constant $\rho_{0k}$ and $v_{0k} = 0$ is considered. This is consistent with \cref{eq:uni} since $\rho_{0k} =  |\psi_{0k}|^2$ and $v_{0k} = \partial_x\theta_{0k}$.
The perturbed state is given by:
\begin{align}\label{perturbation:rho}
  \rho_k &=\rho_{0k}+\epsilon^2\delta\rho_k(\epsilon x, \epsilon t)\\ \label{perturbation:v}
  v_k &=\epsilon^2 \delta v_k(\epsilon x,\epsilon t)
\end{align}
where $\epsilon$ is a small formal parameter. With the perturbation taking this form, \citet{Swarup_etal_2020JPA} showed that a KdV-like model can be derived. 

%\subsection{Spectral analysis}
We begin by studying the dynamics at $\mathcal{O}(\epsilon^0)$
\begin{equation}
  \del_t \hspace{-2pt}
  \begin{pmatrix}
    \delta\rho \\
    \delta v
  \end{pmatrix}\hspace{-4pt} = -\del_x\hspace{1pt}
  \mathcal{A}
  \begin{pmatrix}
    \delta\rho\\
    \delta v
  \end{pmatrix}, \textrm{ where } \mathcal{A} = \left( \begin{array}{cc} \mathbf{0}_{N\times N} & \rho \\ \alpha & \mathbf{0}_{N\times N} \end{array} \right)
\end{equation}
In the above, $\rho$ is $N\times N$ diagonal matrix with elements $\rho_{0k}$. For the base state to be stable, all the eigenvalues of $\mathcal{A}$ have to be real. This is ensured when $h < \min g_j$ \cite{Swarup_etal_2020JPA}. If $\lambda^2$ is an eigenvalue of the matrix $\alpha\rho$, it can be shown that $\pm \lambda$ are eigenvalues of $\mathcal{A}$. Let $\Lambda^2$ be the diagonal matrix with the eigenvalues of the matrix $\alpha\rho$ as its elements. With the corresponding set of eigenvectors denoted as $Q$, the eigenvalue decomposition is $\alpha\rho = Q\Lambda^2Q^{-1}$. The matrix $\mathcal{A}$ can then be decomposed as 
\begin{equation}
\mathcal{A} = V\tilde{\Lambda}V^{-1}
\end{equation}
with
\begin{align}
 \tilde{\Lambda}&=\left(\begin{array}{cc} \Lambda & 0 \\ 0 & -\Lambda \end{array} \right) \textrm{ and } V = \left( \begin{array}{cc} \rho Q\Lambda^{-1} & -\rho Q\Lambda^{-1} \\ Q & Q \end{array}\right) \label{eq:EVD_A}
\end{align}

%\subsection{Nonlinear evolution of perturbations}
The full nonlinear evolution equation for the perturbation is governed by:
\begin{align}
  \left(\partial_T+\mathcal A\partial_X\right)\begin{pmatrix} \delta \rho \\ \delta v \end{pmatrix} &= -\epsilon^2 \partial_X \begin{pmatrix} \mathcal N_1(\delta\rho,\delta v) \\ \mathcal N_{2}(\delta\rho,\delta v,\epsilon)\end{pmatrix} \label{eq:nlin_pevol}
\end{align}
where $X=\epsilon x, T = \epsilon t$ and
\begin{align}
  &\left(\mathcal{N}_1\right)_k = \delta\rho_k\delta v_k \\ 
  &\left(\mathcal N_2\right)_k(\delta \rho,\delta v,\epsilon^2) =\frac{1}{2}\left[ \delta v_k^2 - \frac{2(\rho_{0k}+\epsilon^2\delta\rho_k)\partial_X^2\delta\rho_k - \epsilon^2\partial_X\delta\rho_k^2}{4(\rho_{0k}+\epsilon^2\delta \rho_k)^2}\right]
\end{align}
%With the complete spectral characterisation of $\mathcal{A}$, we now have the means to tackle the inhomogeneous linear problem given in equation \eqref{eq:nlin_pevol}.

We proceed with the perturbation expansion of the fields up to  $\epsilon^2$ as follows
\begin{equation}
  \begin{pmatrix} \delta \rho \\ \delta v \end{pmatrix} = \begin{pmatrix}\delta\rho^{(0)} \\ \delta v^{(0)}\end{pmatrix}  + \epsilon^2 \begin{pmatrix}\delta\rho^{(1)} \\ \delta v^{(1)}\end{pmatrix}
\end{equation}
The general form of the leading order solution is then given by
\begin{equation}
  \begin{pmatrix} \delta \rho^{(0)} \\ \delta v^{(0)} \end{pmatrix} = \sum_{j=1}^{2N}f^{(0)}_j(X-\tilde\lambda_jT,\tau)Ve_j \label{eq:sol_form}
\end{equation}
In the above, $e_j$ is the $j$th column of the $2N \times 2N$ identity matrix, and consequently $Ve_j$ is the $j$th column of $V$. As is commonly seen when using the method of multiple scales, we specify that $f^{(0)}_j$ additionally evolves on a slower time scale $\tau = \epsilon^2 T = \epsilon^3 t$. The equation governing the evolution of $f^{(0)}_j$ is obtained by enforcing the solvability condition on the right-hand side of the equation governing the higher order $\mathcal{O}\left(\epsilon^2\right)$ perturbation:
\begin{align} \label{eq:kdv_basic}
  \partial_\tau f^{(0)}_j +  \left\langle \left(V^{-1}\right)^Te_j, \partial_X \begin{pmatrix} \mathcal N_1\left(\delta \rho^{(0)},\delta v^{(0)}\right) \\ \mathcal N_2\left(\delta \rho^{(0)},\delta v^{(0)},0\right) \end{pmatrix}\right\rangle=0
\end{align}

Equation \eqref{eq:kdv_basic} is a system of $2N$ KdV equations. To derive the KdV system, it is useful to note that we require specific columns of $V$ and $\left(V^{-1}\right)^T$. While $V$ is given by equation \eqref{eq:EVD_A}, $(V^{-1})^T$ can be expressed in terms of $\rho$, $Q$ and $\Lambda$ as:
\begin{align}
  &\left(V^{-1}\right)^T = \frac{1}{2}\left[\begin{array}{cc} QL^{-1} \Lambda & -QL^{-1}\Lambda \\ \rho QL^{-1} & \rho QL^{-1} \end{array}\right]~ \label{eq:V_VinvT}
\end{align}
In the above, $L = Q^T\rho Q$.

\subsection{Derivation of the coupled KdV system}
\label{sec:cKdV_der}

If we consider the initial perturbation to the uniform state to be localized in space, then this perturbation gets resolved into the components given by the terms in the summation on the right-hand side of \cref{eq:sol_form}. Note that each $f_j$ in \cref{eq:sol_form} depends on a spatial coordinate $X-\tilde\lambda_j T$. Thus to an approximation, if $\lambda_j\neq\lambda_j$, then at longer times the $j-$the component $f_j$ is eventually far removed from the $k-$th component $f_k$. In other words, when the eigenvalues are distinct, the initial perturbation evolves into $2N$ distinct non-interacting fields $f_j$ ($N$ propagating to the left; $N$ propagating to the right) each of which is governed by a scalar KdV equation given by \cref{eq:kdv_basic}.

The above situation was precisely the one considered in \citet{Swarup_etal_2020JPA} and \citet{Nimiwal_etal_2021JPA}, namely when the eigenvalues do not repeat. Note that when the eigenvalues $\tilde\lambda_j$ repeat, then even to the lowest approximation, some fields $f_j$ continue to interact even at long times leading to remarkable consequences. In such a situation, the naive expectation is that one then obtains a coupled system of KdV equations. In what follows, we show this is precisely the case.

\citet{Nimiwal_etal_2021JPA} showed that, for the eigenvalue $\pm \lambda_{*}$ to have multiplicity $m$, one requires $\hat{m}~(\equiv m + 1)$ pairs of $(\rho_{0j},g_j)$ to satisfy:
\begin{align}
  \lambda_{*}^{2} = \rho_{0j_1}\left(g_{j_1} - h\right) = \rho_{0j_2}\left(g_{j_2} - h\right) = \dots = \rho_{0j_{\hat{m}}}\left(g_{j_{\hat{m}}} - h\right)\label{eq:r_eval_N_gen}
\end{align}
For convenience, let all the $\rho_{0j}$s that satisfy equation \eqref{eq:r_eval_N_gen} form a continuous segment of the diagonal in $\rho$; likewise, the elements of $\alpha$ will also be consistently placed. Their positions along the diagonal are identified as $j_k$ (with $k = 1,2,\dots,\hat{m}$), and they are identified as $\rho_{0j_k}$. Without loss of generality, we can let $j_k = k$. In addition, we consider here the general scenario where the number of condensates $N$ is greater than $\hat{m}$. 

The density of the $\hat{m}$-th condensate is selected as the reference density $\rho_{0}$. We then have the following expressions for normalised densities:
\begin{align}
  w_j = \frac{\rho_{0j}}{\rho_{0\hat{m}}} = \frac{\rho_{0j}}{\rho_{0}},~\textrm{ for } j = 1,2,\dots,\hat{m}
\end{align}
Note that $w_{\hat{m}} = 1$. Consequently, we can rewrite equation \eqref{eq:r_eval_N_gen} as:
\begin{align}
  \lambda_{*}^{2} = \rho_{0}w_1\left(g_{1} - h\right) =  \dots = \rho_{0}w_m\left(g_{m} - h\right) = \rho_{0}\left(g_{\hat{m}} - h\right) \label{eq:r_eval_N_gen_alt}
\end{align}

The matrix $\alpha\rho - \lambda_{*}^2$ has $\hat{m}$ parallel columns; these columns are also identified by the same $k$ (with $k = 1,2,\dots,\hat{m}$). Every element of the $k$th column of $\alpha\rho - \lambda_{*}^2$ equals $\rho_{0} w_k h$. The eigenvectors of $\alpha\rho$ corresponding to the eigenvalue $\lambda_{*}^2$ are given by $q_{*}^{(k)}$. For $k=1,\dots,m$, the $j$th component of the eigenvector $q_{*}^{(k)}$ is:
\begin{align}
  \left(q_{*}^{(k)}\right)_j =
  \begin{cases}
    w_k, & j = \hat{m}, \\
    -1, \quad\quad\quad & j = k, \\
    0, & \textrm{else.}
  \end{cases}
  \label{eq:reval_evec_N_gen}
\end{align}
With this, we have the columns for $V$ corresponding to the repeated eigenvalues. For $\left(V^{-1}\right)^T$, we need to evaluate $L = Q^T\rho Q$. We do not a priori know the other eigenvalues of $\mathcal{A}$ and the corresponding eigenvectors. Despite this, we can proceed. 
Apart from the first $\hat{m}$ components of the eigenvectors $q_{*}^{(k)}$, the rest are guaranteed to be zero. As a result, there is a $m \times m$ block matrix $\tilde{L}$ embedded within $L$ having the form
\begin{align}
  \tilde{L}_{jk} = \rho_0 w_j w_k + \rho_0 w_j\delta_{jk}
\end{align}
$\tilde{L}$ occupies the rows and the columns numbered from $1$ to $m$ in $L$. It can be verified that its inverse $\tilde{L}^{-1}$ is given by:
\begin{align}
  &\left(\tilde{L}^{-1}\right)_{jk} = -\rho_0^{-1}s(\bar w) + \dfrac{\rho_0^{-1}}{w_j}\delta_{jk}\quad
  \textrm{where } & s(\bar w) = \left(1 + \sum_{l=1}^{m}w_l\right)^{-1}
\end{align}

\noindent $QL^{-1}$ appears in the expression for $\left(V^{-1}\right)^T$ (see equation \eqref{eq:V_VinvT}). As before, we can again exploit the fact that the last $N - \hat{m}$ components of the eigenvectors $q_{*}^{(k)}$ are zero. We can then explicitly write out the elements of the first $m$ columns of $QL^{-1}$. The $k$th column (with $k = 1,2,\dots, m$) of $QL^{-1}$ is given by
\begin{align}
  \left(QL^{-1}\right)_{jk} = 
  \begin{cases}
    \rho_0^{-1}s(\bar w)-\dfrac{\rho_0^{-1}}{w_j}\delta_{jk}, & j = 1,\dots,\hat{m}, % {\color{red}\longleftarrow\; m \mbox{ or }\hat m?}
    \\
    0, & \textrm{ else}.
  \end{cases}
\end{align}
As we now know the form of these specific columns of $QL^{-1}$, the ones we need from $\left(V^{-1}\right)^T$ to get the system of KdV equations are readily available (see equation \eqref{eq:kdv_basic}).

The zeroth order solution in which only the repeated sound speed is accounted for is given by:
\begin{align}
  \left[\begin{array}{c} \delta \rho^{(0)} \\ \delta v^{(0)} \end{array}\right] = \sum_{j = 1}^{m} f_j^{(0)}\left(X - \lambda_{*}T,\tau\right)Ve_j
\end{align}
The KdV equations derived for other sound speeds are not coupled to this system. Therefore initial conditions that are combinations of eigenvectors corresponding to other non-repeating eigenvalues can be considered separately \citep{Swarup_etal_2020JPA}. The elements of $\delta \rho^{(0)}$ and $\delta v^{(0)}$ are then given by:
\begin{align}
  \left(\delta \rho^{(0)}\right)_j =
  \begin{cases}
    \lambda_{*}^{-1}\rho_{0}\sum_{k = 1}^{m}w_kf_{k}^{(0)}, & j = \hat{m}, \\
    -\lambda_{*}^{-1}\rho_{0}w_{j}f_{j}^{(0)}, & j = l \textrm{ for } l = 1,2,\dots,m, \\
    0, & \textrm{else}.
  \end{cases} \label{eq:ZOS_rho_Ngen}\\
  \left(\delta v^{(0)}\right)_j =
  \begin{cases}
    \sum_{k = 1}^{m}w_{k}f_k^{(0)}, & j = \hat{m} \\
    -f_j^{(0)}, & j = l \textrm{ for } l = 1,2,\dots,m, \\
    0, & \textrm{else}. 
  \end{cases} \label{eq:ZOS_vel_Ngen}
\end{align}
Note that the perturbation is non-zero only for those condensates whose properties satisfy equation \eqref{eq:r_eval_N_gen}. Putting all these expressions together in equation \eqref{eq:kdv_basic}, we end up with the following system of KdV equations:
\begin{align}
  \partial_{\tau}f_j^{(0)} +& \frac{3s(\bar w)}{4}\partial_{\xi}\left[\left(\sum_{k = 1}^{m}w_{k}f_k^{(0)}\right)^2 + \sum_{k = 1}^{m}w_{k}\left(f_k^{(0)}\right)^2\right]  \nonumber
  \\ &\quad\quad- \frac{3}{4w_{j}^2}\partial_{\xi} \left(w_{j}f_j^{(0)}\right)^2 - \frac{1}{8\lambda_{*}w_{j}}\partial_{\xi}^3\left(w_{j}f_j^{(0)}\right) = 0 \label{eq:KdV_gen_m}
\end{align}
In the above, $j = 1,2,\cdots,m$ and $\xi = X - \lambda_*T$. With $L_0$ an arbitrary constant, we can rescale the above equations by defining:
\begin{align}
  &\xi = L_0\tilde{\xi}, \tau = -8\lambda_{*}L_0^3\tilde{\tau}, f_{j}^{(0)} = \frac{1}{2\lambda_{*}L_0^2}u_j
\end{align}
Then equation \eqref{eq:KdV_gen_m} can be rewritten as:
\begin{align}
  &\partial_{\tilde{\tau}}u_j + 6u_j\partial_{\tilde{\xi}}u_j + \partial_{\tilde{\xi}}^3u_j = 3s(\bar w)
  \partial_{\tilde{\xi}}\left(\sum_{k,l = 1}^{m}(w_kw_l + w_k\delta_{kl})u_ku_l\right)\: \label{eq:cKdV_v2}
\end{align}
Thus we end with a system of coupled KdV-like fields that describe the dynamics of small-amplitude perturbations to a uniform steady-state solution to \cref{eq:mnls}. We end this section with a few remarks.
\begin{enumerate}
    \item \Cref{eq:cKdV_v2} is precisely of the form discussed in \Cref{sec:mkdv_main}, see \cref{eqn:Nuniversal}, where $s_1(\bar w) = s_2(\bar w)= (1+\sum_{k=1}^m w_k)^{-1}$ and $w_j = \rho_{0j}/\rho_{0\hat m}$.
    \item Only the properties of those condensates which satisfy the necessary condition for eigenvalue multiplicity (equation \eqref{eq:r_eval_N_gen}) determine the coefficients of the system of equations above.
    \item If we are specifically interested only in the dynamics of profiles associated with the repeated sound speed, we might as well disregard the existence of the condensates that do not satisfy equation \eqref{eq:r_eval_N_gen}. The $N$-component nonlinear Schrodinger equation setup acts as though there are only these $\hat{m}$ condensates.
    \item The only substantive feature of the uniform state solution to MNLS that appears in the KdV equations are the densities. Moreover the uniform state densities appear in a symmetric manner: no particular field is distinguished.
    \item The form of $s(\bar w)$ appearing in \cref{eq:cKdV_v2} can be determined can also be determined directly from \crefrange{eqn:Nuniversal}{eqn:Runiversal} by setting equal all symmetric functions of $\bar w$ that appear in \crefrange{eqn:Nuniversal}{eqn:Runiversal}. Indeed
    \begin{align}
    %\[
    s_1(\bar w) = s_2(\bar w) =  1- s_2(\bar w)\sum_j w_j \Rightarrow s_1(\bar w)=s_2(\bar w) = \left(1+\sum_j w_j \right)^{-1}%\]
    \end{align}
    
    \item Every eigenvalue that has a multiplicity (degeneracy), say $m$, is mapped to a system of $m$ coupled KdV equations. Therefore,  MNLS  in its entirety can be described by subsystems of multi-component KdV equations. Of course, for those eigenvalues which have multiplicity of one (distinct/non-degenerate), the mKdV subsystem consists of a single KdV equation.
\end{enumerate}

% \hl{VV: Not sure if any of the following is needed}
% A desirable feature of the above form is that the evolution of all the fields is governed by a single equation. 
% %It is to be also noted that the nonlinear coupling between the different fields is identical in all the equations of the system. 
% Only the properties of those condensates which satisfy the necessary condition for eigenvalue multiplicity (equation \eqref{eq:r_eval_N_gen}) determine the coefficients of the system of equations above. This is a recurring theme throughout this section starting with the expression for the eigenvectors in equation \eqref{eq:reval_evec_N_gen}. Under these circumstances, the system acts as though there are only these $\hat{m}$ condensates. Dynamics pertaining to other condensates can be non-trivial only when additional distinct sound speeds come into play. We note that when $N = m + 1$, there is a slightly more straightforward procedure to yield the same system of $m$ coupled KdV equations (see appendix A). If we are specifically interested only in the dynamics of profiles associated with the repeated sound speed, we might as well disregard the existence of the condensates that do not satisfy equation \eqref{eq:r_eval_N_gen}.

\section{Conclusions and Outlook}
\label{sec:conc}

In this paper, we have established a systematic framework for arriving at multi-component KdV (mKdV) equations from a broader mathematical structure under general and physically motivated assumptions. The obtained mKdV has a rich symmetric structure and is shown to be an effective theory for multi-component systems. More precisely, we demonstrate how $N$ species multi-component nonlinear Schrödinger equations, under appropriate scaling and perturbative treatments, reduce to mKdV equations, provided the system operates within a suitable parameter regime characterized by repeated sound speeds. 

More broadly, our approach offers a unified perspective for understanding the emergence of multi-component partial differential equations from a fundamental mathematical structure. The approach is highly versatile and adaptable for generic multi-component systems. For example, one can adapt the same type of approach given in section~\ref{sec:mkdv_main} for a system of coupled multi-component nonlinear Schrödinger equations. More precisely, on can consider an equation of the following form 
\begin{align}
\partial_t \frac{\delta \mathcal P}{\delta \psi_j} = i \frac{\delta \mathcal H}{\delta \psi^*_j}\label{eqn:conc}
\end{align}
where $\mathcal P$ and $\mathcal H$ are built out of complex-valued fields analogous to what was done in \cref{eqn:PH_m}.  Then there is a  systematic procedure to arrive at the  multi-component nonlinear Schrödinger equations merely by employing the ideas of section~\ref{sec:mkdv_main}. 
Note the adoption of the relevant Poisson bracket to define either the multi-component KdV (in \cref{sec:mkdv_main}) or the multi-component NLS stated above in \cref{eqn:conc}. The details of the derivation of the universal form, as given in \cref{subsec:universal}, remain the same.

It is also worth noting that the mapping presented in section~\ref{sec:mkdv} preserves the all-to-all coupling structure of the equations. Note that \cref{eq:mnls} is indeed of the all-to-all and equal coupling form as is the effective chiral system given in \cref{eq:cKdV_v2}. The preservation of this structure under the scaling and perturbation there is not entirely obvious. It will be interesting to investigate the general criteria for preservation of symmetries.

\section*{Acknowledgements}
We acknowledge the Department of Atomic Energy, Government of India, for their support under Project No. RTI4001. M. K. thanks the VAJRA faculty scheme (No. VJR/2019/000079) from the Science and Engineering Research Board (SERB), Department of Science and Technology, Government of India.

\bibliographystyle{elsarticle-num-names}
\bibliography{references}

\begin{thebibliography}{44}
\expandafter\ifx\csname natexlab\endcsname\relax\def\natexlab#1{#1}\fi
\providecommand{\url}[1]{\texttt{#1}}
\providecommand{\href}[2]{#2}
\providecommand{\path}[1]{#1}
\providecommand{\DOIprefix}{doi:}
\providecommand{\ArXivprefix}{arXiv:}
\providecommand{\URLprefix}{URL: }
\providecommand{\Pubmedprefix}{pmid:}
\providecommand{\doi}[1]{\href{http://dx.doi.org/#1}{\path{#1}}}
\providecommand{\Pubmed}[1]{\href{pmid:#1}{\path{#1}}}
\providecommand{\bibinfo}[2]{#2}
\ifx\xfnm\relax \def\xfnm[#1]{\unskip,\space#1}\fi
%Type = Book
\bibitem[{Ablowitz and Segur(1981)}]{Ablowitz_Segur_1981}
\bibinfo{author}{M.~J. Ablowitz}, \bibinfo{author}{H.~Segur},
  \bibinfo{title}{{Solitons and the Inverse Scattering Transform}},
  \bibinfo{publisher}{Society for Industrial and Applied Mathematics},
  \bibinfo{year}{1981}. \DOIprefix\doi{10.1137/1.9781611970883}.
%Type = Book
\bibitem[{Novikov et~al.(1984)Novikov, Manakov, Pitaevskii, and
  Zakharov}]{Novikov_etal_1984}
\bibinfo{author}{S.~Novikov}, \bibinfo{author}{S.~V. Manakov},
  \bibinfo{author}{L.~P. Pitaevskii}, \bibinfo{author}{V.~E. Zakharov},
  \bibinfo{title}{{Theory of Solitons: The Inverse Scattering Method}},
  \bibinfo{publisher}{Springer New York, NY}, \bibinfo{year}{1984}. \URLprefix
  \url{https://link.springer.com/book/9780306109775}.
%Type = Book
\bibitem[{Ablowitz(2011)}]{Ablowitz_2011}
\bibinfo{author}{M.~J. Ablowitz}, \bibinfo{title}{{Nonlinear Dispersive Waves:
  Asymptotic Analysis and Solitons}}, \bibinfo{publisher}{Cambridge University
  Press}, \bibinfo{year}{2011}. \DOIprefix\doi{10.1017/CBO9780511998324}.
%Type = Article
\bibitem[{van Wijngaarden(1968)}]{Wijngaarden_1968JFM}
\bibinfo{author}{L.~van Wijngaarden},
\newblock \bibinfo{title}{{On the equations of motion for mixtures of liquid
  and gas bubbles}},
\newblock \bibinfo{journal}{J.~Fluid.~Mech.} \bibinfo{volume}{33}
  (\bibinfo{year}{1968}) \bibinfo{pages}{465--474}.
  \DOIprefix\doi{10.1017/S002211206800145X}.
%Type = Article
\bibitem[{Hammack and Segur(1974)}]{Hammack_Segur_1974JFM}
\bibinfo{author}{J.~L. Hammack}, \bibinfo{author}{H.~Segur},
\newblock \bibinfo{title}{{The Korteweg--de Vries equation and water waves.
  Part 2. Comparison with experiments}},
\newblock \bibinfo{journal}{J.~Fluid.~Mech.} \bibinfo{volume}{65}
  (\bibinfo{year}{1974}) \bibinfo{pages}{289--314}.
  \DOIprefix\doi{10.1017/S002211207400139X}.
%Type = Article
\bibitem[{Zabusky and Kruskal(1965)}]{Zabusky_Kruskal_1965PRL}
\bibinfo{author}{N.~J. Zabusky}, \bibinfo{author}{M.~D. Kruskal},
\newblock \bibinfo{title}{{Interaction of ``solitons'' in a collisionless
  plasma and the recurrence of initial states}},
\newblock \bibinfo{journal}{Phys.~Rev.~Lett.} \bibinfo{volume}{15}
  (\bibinfo{year}{1965}) \bibinfo{pages}{240}.
  \DOIprefix\doi{10.1103/PhysRevLett.15.240}.
%Type = Article
\bibitem[{Kivshar(1993)}]{Kivshar_1993JQE}
\bibinfo{author}{Y.~S. Kivshar},
\newblock \bibinfo{title}{{Dark Solitons in Nonlinear Optics}},
\newblock \bibinfo{journal}{IEEE~J.~Quantum~Electron.} \bibinfo{volume}{29}
  (\bibinfo{year}{1993}) \bibinfo{pages}{250--264}.
  \DOIprefix\doi{10.1109/3.199266}.
%Type = Article
\bibitem[{Miura et~al.(1968)Miura, Gardner, and Kruskal}]{Miuraetal_1968JMP}
\bibinfo{author}{R.~M. Miura}, \bibinfo{author}{C.~S. Gardner},
  \bibinfo{author}{M.~D. Kruskal},
\newblock \bibinfo{title}{{Korteweg--de Vries Equation and Generalizations. II.
  Existence of Conservation Laws and Constants of Motion}},
\newblock \bibinfo{journal}{J.~Math.~Phys.} \bibinfo{volume}{9}
  (\bibinfo{year}{1968}) \bibinfo{pages}{1204--1209}.
  \DOIprefix\doi{10.1063/1.1664701}.
%Type = Article
\bibitem[{Miura(1976)}]{Miura_1976SIAMRev}
\bibinfo{author}{R.~M. Miura},
\newblock \bibinfo{title}{{The Korteweg--deVries Equation: A Survey of
  Results}},
\newblock \bibinfo{journal}{SIAM Rev.} \bibinfo{volume}{18}
  (\bibinfo{year}{1976}) \bibinfo{pages}{412--459}.
  \DOIprefix\doi{10.1137/1018076}.
%Type = Article
\bibitem[{Gardner(1971)}]{Gardner_1971JMP}
\bibinfo{author}{C.~S. Gardner},
\newblock \bibinfo{title}{{{K}orteweg--de{V}ries {E}quation and
  {G}eneralizations. {IV}. The Korteweg--de Vries Equation as a Hamiltonian
  System}},
\newblock \bibinfo{journal}{J.~Math.~Phys.} \bibinfo{volume}{12}
  (\bibinfo{year}{1971}) \bibinfo{pages}{1548--1551}.
  \DOIprefix\doi{10.1063/1.1665772}.
%Type = Article
\bibitem[{Zakharov and Faddeev(1971)}]{Zakharov_Faddeev_1971FAA}
\bibinfo{author}{V.~E. Zakharov}, \bibinfo{author}{L.~D. Faddeev},
\newblock \bibinfo{title}{{Korteweg--de Vries equation: A completely integrable
  Hamiltonian system}},
\newblock \bibinfo{journal}{Funct.~Anal.~Its~Appl.} \bibinfo{volume}{5}
  (\bibinfo{year}{1971}) \bibinfo{pages}{280--287}.
  \DOIprefix\doi{10.1007/BF01086739}.
%Type = Article
\bibitem[{Hirota(1971)}]{Hirota_1971PRL}
\bibinfo{author}{R.~Hirota},
\newblock \bibinfo{title}{{Exact Solution of the Korteweg--de Vries Equation
  for Multiple Collisions of Solitons}},
\newblock \bibinfo{journal}{Phys.~Rev.~Lett.} \bibinfo{volume}{27}
  (\bibinfo{year}{1971}) \bibinfo{pages}{1192}.
  \DOIprefix\doi{10.1103/PhysRevLett.27.1192}.
%Type = Book
\bibitem[{Hirota(2004)}]{Hirota_2004}
\bibinfo{author}{R.~Hirota}, \bibinfo{title}{{The {D}irect {M}ethod in
  {S}oliton {T}heory}}, \bibinfo{publisher}{Cambridge University Press},
  \bibinfo{year}{2004}. \DOIprefix\doi{10.1017/CBO9780511543043}.
%Type = Article
\bibitem[{Gear and Grimshaw(1984)}]{Gear_Grimshaw_1984SAM}
\bibinfo{author}{J.~A. Gear}, \bibinfo{author}{R.~Grimshaw},
\newblock \bibinfo{title}{{Weak and Strong Interactions between Internal
  Solitary Waves}},
\newblock \bibinfo{journal}{Stud.~Appl.~Math.} \bibinfo{volume}{70}
  (\bibinfo{year}{1984}) \bibinfo{pages}{235--258}.
  \DOIprefix\doi{10.1002/sapm1984703235}.
%Type = Article
\bibitem[{Majda and Biello(2003)}]{Majda_Biello_2003JAS}
\bibinfo{author}{A.~J. Majda}, \bibinfo{author}{J.~A. Biello},
\newblock \bibinfo{title}{{The Nonlinear Interaction of Barotropic and
  Equatorial Baroclinic Rossby Waves}},
\newblock \bibinfo{journal}{J.~Atmos.~Sci.} \bibinfo{volume}{60}
  (\bibinfo{year}{2003}) \bibinfo{pages}{1809--1821}.
  \DOIprefix\doi{10.1175/1520-0469(2003)060<1809:TNIOBA>2.0.CO;2}.
%Type = Article
\bibitem[{Lou et~al.(2006)Lou, Tong, Hu, and Tang}]{Lou_etal_2006JPAMG}
\bibinfo{author}{S.~Y. Lou}, \bibinfo{author}{B.~Tong}, \bibinfo{author}{H.-c.
  Hu}, \bibinfo{author}{X.-y. Tang},
\newblock \bibinfo{title}{{Coupled {KdV} equations derived from two-layer
  fluids}},
\newblock \bibinfo{journal}{J.~Phys.~A:~Math.~Gen.} \bibinfo{volume}{39}
  (\bibinfo{year}{2006}) \bibinfo{pages}{513}.
  \DOIprefix\doi{10.1088/0305-4470/39/3/005}.
%Type = Article
\bibitem[{Brazhnyi and Konotop(2005)}]{Brazhnyi_Konotop_2005PRE}
\bibinfo{author}{V.~A. Brazhnyi}, \bibinfo{author}{V.~V. Konotop},
\newblock \bibinfo{title}{{Stable and unstable vector dark solitons of coupled
  nonlinear {S}chr{\"o}dinger equations: {A}pplication to two-component
  {B}ose-{E}instein condensates}},
\newblock \bibinfo{journal}{Phys.~Rev.~E} \bibinfo{volume}{72}
  (\bibinfo{year}{2005}) \bibinfo{pages}{026616}.
  \DOIprefix\doi{10.1103/PhysRevE.72.026616}.
%Type = Article
\bibitem[{Grimshaw and Malomed(1994)}]{Grimshaw_Malomed_1994PRL}
\bibinfo{author}{R.~Grimshaw}, \bibinfo{author}{B.~A. Malomed},
\newblock \bibinfo{title}{{New Type of Gap Soliton in a Coupled Korteweg--de
  Vries Wave System}},
\newblock \bibinfo{journal}{Phys.~Rev.~Lett.} \bibinfo{volume}{72}
  (\bibinfo{year}{1994}) \bibinfo{pages}{949}.
  \DOIprefix\doi{10.1103/PhysRevLett.72.949}.
%Type = Article
\bibitem[{Champneys et~al.(2000)Champneys, Groves, and
  Woods}]{Champneys_etal_2000PLA}
\bibinfo{author}{A.~R. Champneys}, \bibinfo{author}{M.~D. Groves},
  \bibinfo{author}{P.~D. Woods},
\newblock \bibinfo{title}{{A global characterization of gap solitary-wave
  solutions to a coupled {KdV} system}},
\newblock \bibinfo{journal}{Phys.~Lett.~A} \bibinfo{volume}{271}
  (\bibinfo{year}{2000}) \bibinfo{pages}{178--190}.
  \DOIprefix\doi{10.1016/S0375-9601(00)00355-8}.
%Type = Article
\bibitem[{Grimshaw and Iooss(2003)}]{Grimshaw_Iooss_2003MCS}
\bibinfo{author}{R.~Grimshaw}, \bibinfo{author}{G.~Iooss},
\newblock \bibinfo{title}{{Solitary waves of a coupled {Korteweg-de Vries}
  system}},
\newblock \bibinfo{journal}{Math.~Comput.~Simulation} \bibinfo{volume}{62}
  (\bibinfo{year}{2003}) \bibinfo{pages}{31--40}.
  \DOIprefix\doi{10.1016/S0378-4754(02)00189-1}.
%Type = Article
\bibitem[{Fochesato et~al.(2005)Fochesato, Dias, and
  Grimshaw}]{Fochesato_etal_2005PD}
\bibinfo{author}{C.~Fochesato}, \bibinfo{author}{F.~Dias},
  \bibinfo{author}{R.~Grimshaw},
\newblock \bibinfo{title}{{Generalized solitary waves and fronts in coupled
  {K}orteweg--de {V}ries systems}},
\newblock \bibinfo{journal}{Physica D} \bibinfo{volume}{210}
  (\bibinfo{year}{2005}) \bibinfo{pages}{96--117}.
  \DOIprefix\doi{10.1016/j.physd.2005.07.010}.
%Type = Article
\bibitem[{Hirota and Satsuma(1981)}]{Hirota_Satsuma_1981PL}
\bibinfo{author}{R.~Hirota}, \bibinfo{author}{J.~Satsuma},
\newblock \bibinfo{title}{{Soliton solutions of a coupled Korteweg--de Vries
  equation}},
\newblock \bibinfo{journal}{Phys.~Lett.~A} \bibinfo{volume}{85}
  (\bibinfo{year}{1981}) \bibinfo{pages}{407--408}.
  \DOIprefix\doi{10.1016/0375-9601(81)90423-0}.
%Type = Article
\bibitem[{Satsuma and Hirota(1982)}]{Satsuma_Hirota_1982JPSJ}
\bibinfo{author}{J.~Satsuma}, \bibinfo{author}{R.~Hirota},
\newblock \bibinfo{title}{{A Coupled KdV Equation is One Case of the
  Four-Reduction of the KP Hierarchy}},
\newblock \bibinfo{journal}{J.~Phys.~Soc.~Jpn.} \bibinfo{volume}{51}
  (\bibinfo{year}{1982}) \bibinfo{pages}{3390--3397}.
  \DOIprefix\doi{10.1143/JPSJ.51.3390}.
%Type = Article
\bibitem[{Ito(1982)}]{Ito_1982PL}
\bibinfo{author}{M.~Ito},
\newblock \bibinfo{title}{{Symmetries and conservation laws of a coupled
  nonlinear wave equation}},
\newblock \bibinfo{journal}{Phys.~Lett.~A} \bibinfo{volume}{91}
  (\bibinfo{year}{1982}) \bibinfo{pages}{335--338}.
  \DOIprefix\doi{10.1016/0375-9601(82)90426-1}.
%Type = Article
\bibitem[{Fuchssteiner(1982)}]{Fuchssteiner_1982PTP}
\bibinfo{author}{B.~Fuchssteiner},
\newblock \bibinfo{title}{{The Lie Algebra Structure of Degenerate Hamiltonian
  and Bi-Hamiltonian Systems}},
\newblock \bibinfo{journal}{Prog.~Theor.~Phys.} \bibinfo{volume}{68}
  (\bibinfo{year}{1982}) \bibinfo{pages}{1082--1104}.
  \DOIprefix\doi{10.1143/PTP.68.1082}.
%Type = Article
\bibitem[{Baker et~al.(1995)Baker, Enolskii, and Fordy}]{Baker_etal_1995PLA}
\bibinfo{author}{S.~Baker}, \bibinfo{author}{V.~Z. Enolskii},
  \bibinfo{author}{A.~P. Fordy},
\newblock \bibinfo{title}{{Integrable quartic potentials and coupled {KdV}
  equations}},
\newblock \bibinfo{journal}{Phys.~Lett.~A} \bibinfo{volume}{201}
  (\bibinfo{year}{1995}) \bibinfo{pages}{167--174}.
  \DOIprefix\doi{10.1016/0375-9601(95)00267-7}.
%Type = Article
\bibitem[{Karasu(1997)}]{Karasu_1997JMP}
\bibinfo{author}{A.~K. Karasu},
\newblock \bibinfo{title}{{Painlev{\'e} classification of coupled Korteweg--de
  Vries systems}},
\newblock \bibinfo{journal}{J.~Math.~Phys.} \bibinfo{volume}{38}
  (\bibinfo{year}{1997}) \bibinfo{pages}{3616--3622}.
  \DOIprefix\doi{10.1063/1.532056}.
%Type = Article
\bibitem[{Sakovich(1999)}]{Sakovich_1999JNMP}
\bibinfo{author}{S.~Y. Sakovich},
\newblock \bibinfo{title}{{Coupled KdV Equations of Hirota--Satsuma Type}},
\newblock \bibinfo{journal}{J.~Nonlinear~Math.~Phys.} \bibinfo{volume}{6}
  (\bibinfo{year}{1999}) \bibinfo{pages}{255--262}.
  \DOIprefix\doi{10.2991/jnmp.1999.6.3.2}.
%Type = Article
\bibitem[{Foursov(2000)}]{Foursov_2000IP}
\bibinfo{author}{M.~V. Foursov},
\newblock \bibinfo{title}{{On integrable coupled {KdV}-type systems}},
\newblock \bibinfo{journal}{Inverse~Probl.} \bibinfo{volume}{16}
  (\bibinfo{year}{2000}) \bibinfo{pages}{259}.
  \DOIprefix\doi{10.1088/0266-5611/16/1/319}.
%Type = Article
\bibitem[{Li(2004)}]{Li_2004JPSJ}
\bibinfo{author}{C.-X. Li},
\newblock \bibinfo{title}{{A Hierarchy of Coupled Korteweg--de Vries Equations
  and the Corresponding Finite-Dimensional Integrable System}},
\newblock \bibinfo{journal}{J.~Phys.~Soc.~Jpn.} \bibinfo{volume}{73}
  (\bibinfo{year}{2004}) \bibinfo{pages}{327--331}.
  \DOIprefix\doi{10.1143/JPSJ.73.327}.
%Type = Article
\bibitem[{Wang(2010)}]{Wang_2010AMC}
\bibinfo{author}{D.-S. Wang},
\newblock \bibinfo{title}{{Integrability of a coupled KdV system: Painlev{\'e}
  property, Lax pair and B{\"a}cklund transformation}},
\newblock \bibinfo{journal}{Appl.~Math.~Comput.} \bibinfo{volume}{216}
  (\bibinfo{year}{2010}) \bibinfo{pages}{1349--1354}.
  \DOIprefix\doi{10.1016/j.amc.2010.02.030}.
%Type = Article
\bibitem[{Yan(2003)}]{Yan_2003CSF}
\bibinfo{author}{Z.~Yan},
\newblock \bibinfo{title}{{The extended Jacobian elliptic function expansion
  method and its application in the generalized Hirota--Satsuma coupled KdV
  system}},
\newblock \bibinfo{journal}{Chaos~Soliton.~Fract.} \bibinfo{volume}{15}
  (\bibinfo{year}{2003}) \bibinfo{pages}{575--583}.
  \DOIprefix\doi{10.1016/S0960-0779(02)00145-5}.
%Type = Article
\bibitem[{Ganji and Rafei(2006)}]{Ganji_Rafei_2006PLA}
\bibinfo{author}{D.~D. Ganji}, \bibinfo{author}{M.~Rafei},
\newblock \bibinfo{title}{{Solitary wave solutions for a generalized
  Hirota--Satsuma coupled KdV equation by homotopy perturbation method}},
\newblock \bibinfo{journal}{Phys.~Lett.~A} \bibinfo{volume}{356}
  (\bibinfo{year}{2006}) \bibinfo{pages}{131--137}.
  \DOIprefix\doi{10.1016/j.physleta.2006.03.039}.
%Type = Article
\bibitem[{Hu et~al.(2006)Hu, Tong, and Lou}]{Hu_etal_2006PLA}
\bibinfo{author}{H.~C. Hu}, \bibinfo{author}{B.~Tong}, \bibinfo{author}{S.~Y.
  Lou},
\newblock \bibinfo{title}{{Nonsingular positon and complexiton solutions for
  the coupled {KdV} system}},
\newblock \bibinfo{journal}{Phys.~Lett.~A} \bibinfo{volume}{351}
  (\bibinfo{year}{2006}) \bibinfo{pages}{403--412}.
  \DOIprefix\doi{10.1016/j.physleta.2005.11.047}.
%Type = Article
\bibitem[{Wazwaz(2011)}]{Wazwaz_2011CEJP}
\bibinfo{author}{A.-M. Wazwaz},
\newblock \bibinfo{title}{{Integrability of coupled KdV equations}},
\newblock \bibinfo{journal}{Cent.~Eur.~J.~Phys.} \bibinfo{volume}{9}
  (\bibinfo{year}{2011}) \bibinfo{pages}{835--840}.
  \DOIprefix\doi{10.2478/s11534-010-0084-y}.
%Type = Article
\bibitem[{Svinolupov(1991)}]{Svinolupov_1991}
\bibinfo{author}{S.~I. Svinolupov},
\newblock \bibinfo{title}{{Jordan algebras and generalized Korteweg--de Vries
  equations}},
\newblock \bibinfo{journal}{Theor.~Mat.~Phys.} \bibinfo{volume}{87}
  (\bibinfo{year}{1991}) \bibinfo{pages}{611--620}.
  \DOIprefix\doi{10.1007/BF01017947}.
%Type = Article
\bibitem[{G{\"u}rses and Karasu(1996)}]{Gurses_Karasu_1996PLA}
\bibinfo{author}{M.~G{\"u}rses}, \bibinfo{author}{A.~Karasu},
\newblock \bibinfo{title}{{Degenerate Svinolupov KdV systems}},
\newblock \bibinfo{journal}{Phys.~Lett.~A} \bibinfo{volume}{214}
  (\bibinfo{year}{1996}) \bibinfo{pages}{21--26}.
  \DOIprefix\doi{10.1016/0375-9601(96)00171-5}.
%Type = Article
\bibitem[{Miller and Christiansen(2000)}]{Miller_Christiansen_2000PS}
\bibinfo{author}{P.~D. Miller}, \bibinfo{author}{P.~L. Christiansen},
\newblock \bibinfo{title}{{A Coupled Korteweg-de Vries System and Mass
  Exchanges Among Solitons}},
\newblock \bibinfo{journal}{Phys.~Scripta} \bibinfo{volume}{61}
  (\bibinfo{year}{2000}) \bibinfo{pages}{518}.
  \DOIprefix\doi{10.1238/Physica.Regular.061a00518}.
%Type = Article
\bibitem[{Swarup et~al.(2020)Swarup, Vasan, and Kulkarni}]{Swarup_etal_2020JPA}
\bibinfo{author}{S.~Swarup}, \bibinfo{author}{V.~Vasan},
  \bibinfo{author}{M.~Kulkarni},
\newblock \bibinfo{title}{{Provable bounds for the {K}orteweg--de {V}ries
  reduction in multi-component nonlinear {S}chr{\"o}dinger equation}},
\newblock \bibinfo{journal}{J.~Phys.~A: Math.~Theor.} \bibinfo{volume}{53}
  (\bibinfo{year}{2020}) \bibinfo{pages}{135206}.
  \DOIprefix\doi{10.1088/1751-8121/ab6f19}.
%Type = Article
\bibitem[{Nimiwal et~al.(2021)Nimiwal, Satpathi, Vasan, and
  Kulkarni}]{Nimiwal_etal_2021JPA}
\bibinfo{author}{R.~Nimiwal}, \bibinfo{author}{U.~Satpathi},
  \bibinfo{author}{V.~Vasan}, \bibinfo{author}{M.~Kulkarni},
\newblock \bibinfo{title}{{Soliton-like behaviour in non-integrable systems}},
\newblock \bibinfo{journal}{J.~Phys.~A: Math.~Theor.} \bibinfo{volume}{54}
  (\bibinfo{year}{2021}) \bibinfo{pages}{425701}.
  \DOIprefix\doi{10.1088/1751-8121/ac1ee5}.
%Type = Article
\bibitem[{Inouye et~al.(1998)Inouye, Andrews, Stenger, Miesner, Stamper-Kurn,
  and Ketterle}]{Inouye_etal_1998}
\bibinfo{author}{S.~Inouye}, \bibinfo{author}{M.~R. Andrews},
  \bibinfo{author}{J.~Stenger}, \bibinfo{author}{H.-J. Miesner},
  \bibinfo{author}{D.~M. Stamper-Kurn}, \bibinfo{author}{W.~Ketterle},
\newblock \bibinfo{title}{{Observation of Feshbach resonances in a
  Bose--Einstein condensate}},
\newblock \bibinfo{journal}{Nature} \bibinfo{volume}{392}
  (\bibinfo{year}{1998}) \bibinfo{pages}{151--154}.
  \DOIprefix\doi{10.1038/32354}.
%Type = Article
\bibitem[{Roati et~al.(2007)Roati, Zaccanti, d’Errico, Catani, Modugno,
  Simoni, Inguscio, and Modugno}]{Roati_etal_2007PRL}
\bibinfo{author}{G.~Roati}, \bibinfo{author}{M.~Zaccanti},
  \bibinfo{author}{C.~d’Errico}, \bibinfo{author}{J.~Catani},
  \bibinfo{author}{M.~Modugno}, \bibinfo{author}{A.~Simoni},
  \bibinfo{author}{M.~Inguscio}, \bibinfo{author}{G.~Modugno},
\newblock \bibinfo{title}{{$^{39}$K Bose-Einstein Condensate with Tunable
  Interactions}},
\newblock \bibinfo{journal}{Phys.~Rev.~Lett.} \bibinfo{volume}{99}
  (\bibinfo{year}{2007}) \bibinfo{pages}{010403}.
  \DOIprefix\doi{10.1103/PhysRevLett.99.010403}.
%Type = Article
\bibitem[{Thalhammer et~al.(2008)Thalhammer, Barontini, De~Sarlo, Catani,
  Minardi, and Inguscio}]{Thalhammer_etal_2008PRL}
\bibinfo{author}{G.~Thalhammer}, \bibinfo{author}{G.~Barontini},
  \bibinfo{author}{L.~De~Sarlo}, \bibinfo{author}{J.~Catani},
  \bibinfo{author}{F.~Minardi}, \bibinfo{author}{M.~Inguscio},
\newblock \bibinfo{title}{{Double Species Bose-Einstein Condensate with Tunable
  Interspecies Interactions}},
\newblock \bibinfo{journal}{Phys.~Rev.~Lett.} \bibinfo{volume}{100}
  (\bibinfo{year}{2008}) \bibinfo{pages}{210402}.
  \DOIprefix\doi{10.1103/PhysRevLett.100.210402}.
%Type = Article
\bibitem[{Wacker et~al.(2015)Wacker, J{\o}rgensen, Birkmose, Horchani, Ertmer,
  Klempt, Winter, Sherson, and Arlt}]{Wacker_etal_2015PRA}
\bibinfo{author}{L.~Wacker}, \bibinfo{author}{N.~B. J{\o}rgensen},
  \bibinfo{author}{D.~Birkmose}, \bibinfo{author}{R.~Horchani},
  \bibinfo{author}{W.~Ertmer}, \bibinfo{author}{C.~Klempt},
  \bibinfo{author}{N.~Winter}, \bibinfo{author}{J.~Sherson},
  \bibinfo{author}{J.~J. Arlt},
\newblock \bibinfo{title}{{Tunable dual-species Bose-Einstein condensates of
  $^{39}$K and $^{87}$Rb}},
\newblock \bibinfo{journal}{Phys.~Rev.~A} \bibinfo{volume}{92}
  (\bibinfo{year}{2015}) \bibinfo{pages}{053602}.
  \DOIprefix\doi{10.1103/PhysRevA.92.053602}.

\end{thebibliography}

\end{document}